\documentclass[proof]{WileyASNA-v1}

\usepackage[utf8]{inputenc}
\usepackage{booktabs}
\usepackage{caption}
\usepackage{graphicx}
\usepackage{verbatim}
\usepackage{bm}
\articletype{Article Type}%

\begin{document}

\title{A short review on the pulsar magnetic inclination angles}

\author[1,2,3]{Biao-Peng Li}

\author[1,2]{Zhi-Fu Gao*}

\authormark{Biao-Peng Li}

\address[1]{Xinjiang Astronomical Observatory, CAS,150,
Science 1-Street, Urumqi, Xinjiang, 830011, China}

\address[2]{Key Laboratory of Radio Astronomy, Chinese Academy of Sciences,
West Beijing Road, Nanjing, 210008, China}

\address[3]{University of Chinese Academy of Sciences, No.19, Yuquan Road, Beijing, 100049, China}

\corres{*Zhi-Fu Gao. Xinjiang Astronomical Observatory, CAS,150, Science
1-Street, Urumqi, Xinjiang, 830011, China.\\
\email{zhifugao@xao.ac.cn}}

\abstract{The inclination angle $\chi$ between magnetic and rotation axes of
pulsars is an important parameter in pulsar physics. The changes in the inclination
angle of a pulsar would lead to observable effects, such as changes in the pulse
beam width and braking index of the star. In this paper, we perform a short review
on the evolution of pulsar's magnetic inclination angle, as well as the latest
research progress. Using an alignment rotator model in vacuum, we investigate the
magnetic inclination angle change rates for 12 high-braking index pulsars
without glitch, whose timing observations are obtained using the Nanshan 25-m Radio
Telescope at Xinjiang Astronomical Observatory. For our purpose, three representative pulsars
J0157+6212, J1743-3150 and J1857+0526 are chosen and their rotation and inclination angle
evolutions are further investigated. In the future, radio and X-ray polarimetric observations
will provide more information about the inclination angles of pulsars, which could help
us understand the origin of the variations in $\chi$ of pulsars and shed light on the
range of possibilities of pulsar magnetic field configuration. A continuous study of
the pulsar inclination angle will provide an important window into additional physical
processes at work in the young and highly magnetized pulsars.}


\begin{comment}
\jnlcitation{\cname{%
\author{Williams K.},
\author{B. Hoskins},
\author{R. Lee},
\author{G. Masato}, and
\author{T. Woollings}} (\cyear{2016}),
\ctitle{A regime analysis of Atlantic winter jet variability applied to evaluate HadGEM3-GC2},
\cjournal{Q.J.R. Meteorol. Soc.}, \cvol{2017;00:1--6}.}
\end{comment}
\fundingInfo{Chinese National Science Foundation through grants No.12041304.}

\maketitle

\begin{comment}
\footnotetext{\textbf{Abbreviations:} ANA, anti-nuclear antibodies; APC,
antigen-presenting cells; IRF, interferon regulatory factor}
\end{comment}

\section{Introduction and motivation}\label{sec1}
Pulsars are highly magnetized and rapidly rotating neutron stars with rich emission phenomena,
providing us an opportunity to explore physics under extreme conditions. A simplified model of pulsars is the classic vacuum magnetic dipole radiation where there is an angle between the magnetic axis and the axis of rotation\,(the inclination angle $\chi$), which is an important parameter in pulsar physics. The variations in $\chi$ of a pulsar would lead to observable effects, such as changes in the pulse beam width. The knowledge of initial angle $\chi_{0}$ would provide valuable information about how pulsar magnetic fields are formed. Since the rotational energy loss rate and radio luminosity of pulsars depend on the inclination $\chi$, the values and time development of $\chi$ have been described by various models\, (see e.g.\,\citep{Goldreich1969,ruderman1975ApJ,harding2002ApJ,istomin2007ARep,barsukov2009ARep,kou2019ApJ}).

In general, the value of $\chi$ of a pulsar can be independently determined by modeling of the gamma-ray or radio pulse profiles. Measurements of $\chi$ were used to study possible alignment\,(or counter alignment) or decay of pulsar electromagnetic\,(EM) fields\,\citep{Srinivasan1989}. There are more than 200 pulsars with reported values of $\chi$ by polarization measurements or other ways, and the angles distribute in a wide range of about $5^{\circ}-87^{\circ}$\,(see e.g.\,\citep{Lyne1988MNRAS,Rankin1990ApJ,Malov2011ARep,tedila2022ApJ}). With the exception of the Crab pulsar, it is difficult to accurately measure the variation of other pulsars' inclination angles, which involves many reasons, such as polarization position angle, dispersion measurement, pulse profile change and so on. The Crab pulsar is the only pulsar for which a change in $\chi$ has been measured, at $+(0.62\pm0.03)^{\circ}$ per century by observing the change in its radio pulse profile\,\citep{Lyne2013}. However, the reason for the increase in $\chi$ of this pulsar remains a mystery.

Since pulsars were discovered, to reveal the mysteries of radiation phenomena and rotational irregularity observed in pulsars, some theoretical models on the pulsar inclination angle
have been proposed\,(see e.g.\,\citep{harding2002ApJ,beskin2007ApSS,kou2019ApJ}).
However, there are few reviews on the pulsar inclination angles and their evolutions, which makes it difficult to provide us with a more comprehensive understanding of the evolution mechanism of pulsar inclination angles. The variation of $\chi$ could be associated with some observable phenomena in pulsars, such as  slow glitches\,\citep{Shabanova2005MN}, profile change\,\citep{wen2016AA,wen2021ApJ,wang2020ApJ},
drifting subpulse\,\citep{yuen2014PASA,yuen2019MNRAS,yan2019MNRAS,yan2020MNRAS}, pulsar braking indices that are different from the canonic dipole value\,\citep{Eksi2016,Lander2018,Tong2017}. In view of this, we will perform a short review on the pulsar magnetic inclination angle, as well as their evolutions in this work.

 This paper is organized as follows. In Section 2 we give a brief overview of the origin of variations in $\chi$ of pulsars; 
 In Section 3 we review the trends and forms of variations in $\chi$ of pulsars; In Section 4 we review some observable 
 effects in pulsars possibly related to variations in $\chi$; In Section 5 we investigate the inclination angles and their 
 evolutions for several pulsars observed by the Nanshan 25-m Radio Telescope within the framework of vacuum model. Section 6 
 gives a summary of this work.
\section{The origin of variations in inclination angle}\label{sec2}
The inclination angle $\chi$ of a pulsar cannot remains constant throughout its lifetime. 
The variation in $\chi$ either comes from inside or outside of the star, although variation 
mechanisms have not yet been understood. The angle $\chi$ increases or decreases depending on 
the mechanism by which the pulsar loses energy. 
\subsection{External origin}
\subsubsection{The vacuum model}
If a pulsar's radiant energy is provided entirely by its rotational energy, pulsar rotating 
in vacuum will be slowed down by the radiation torques. \citet{Michel1970} pointed out that 
there are two torques acting on the pulsar in the vacuum model: the spin-down torque that causes 
the rotational angular velocity to decrease with time and the alignment torque causing the 
magnetic inclination to evolve towards aligned configuration,
\begin{eqnarray}
    I\frac{d\Omega}{dt}=-\frac{2}{3}\frac{M^{2}}{c^{3}}\Omega^{3}\sin^{2}{\chi},
\end{eqnarray}
\label{1}
\begin{eqnarray}
    ~~~~~~~~I\frac{d\chi}{dt}=-\frac{2}{3}\frac{M^{2}}{c^{3}}\Omega^{2}\sin{\chi}\cos{\chi},
\label{2}
\end{eqnarray}
where $I$ is the moment of inertia, $M$ the magnetic dipole moment, $\Omega$ the rotational angular 
velocity, and $c$ the speed of light. The alignment torque is an intrinsic component of the torque due to 
the magnetic dipole radiation. The vacuum model does not consider the presence of pulsar magnetosphere 
and is a very simplified pulsar model. Interestingly, due to its simplification, the vacuum model is still 
the most commonly used theoretical model in pulsar study.
\subsubsection{The plasma-filled model}
The pulsar magnetospheres are expected to be filled with a co-rotating plasma, which is formed by charged 
particles stripped from the surface of the star\,\citep{Goldreich1969}, and then accelerated by
rotating-induced electric field along curved magnetic field lines to give an excess of
electron–positron pair discharges, and only this configuration can keep the pulsar
active. A change in the charge density and/or the current density will result in a variation in magnetospheric torque, thus $\chi$ will change accordingly~\citet{Spitkovsky2006}.

\citet{Philippov2014} firstly gave a self-explanatory for the plasma
effects by analysing the results of time-dependent
force-free and magnetohydrodynamic\,(MHD) simulations of pulsar magnetosphere.
Taking account into the plasma effects, the torque equations become
\begin{eqnarray}
    I\frac{d\Omega}{dt}=-\frac{2}{3}\frac{M^{2}}{c^{3}}\Omega^{3}(1+\sin^{2}{\chi}),
  \label{3}
\end{eqnarray}
\begin{eqnarray}
    I\frac{d\chi}{dt}=-\frac{2}{3}\frac{M^{2}}{c^{3}}\Omega^{2}\sin{\chi}\cos{\chi}.
     \label{4}
\end{eqnarray}
Obviously, both Eq.(1) and Eq.(3) have a  comment magnetic dipole braking torque in vacuum,
\begin{eqnarray}
    \bm{K}_{\text{dip}}=-\frac{2}{3} \frac{M^{2}}{c^{3}}\Omega^{2}\bm{\Omega}\sin^{2}{\chi},
  \label{5}
\end{eqnarray}
which results in the loss of angular momentum of the pulsar.
~\citet{Philippov2014} also showed that pulsars evolve so as to minimize their
spin-down luminosities: both the vacuum and plasma-filled pulsars evolve
towards the aligned configuration. However, no pulsars with zero magnetic
inclination angles have been observed, and the increasing in $\chi$ of the Crab pulsar is difficult to be explained alone either
in vacuum model or in plasma-filled model.
\subsubsection{The current-braking model}
Oppositely, a current-braking pulsar will evolve towards the orthogonal
configuration. According to \citet{beskin1993,barsukov2009ARep}, there is a region in the pulsar magnetosphere, called
the pulsar tube, which is formed by
magnetic-field lines crossing the light cylinder. An electric current $\bm{j}$ flows in the pulsar tube,
and the current flowing from the star surface comes back in a narrow layer near
the pulsar tube boundary due to conservation of charge. For these currents to be
closed, a part of their paths must cross the magnetic field $\bm{B}$. Thus,
the Lorentz force $\bm{F}=[\bm{j}\times \bm{B}]/c$ acts on the polar-cap region, the associated
torque acting on the crust is given by
\begin{eqnarray}
    \bm{K}_{\text{cur}}=-\frac{2}{3}\frac{M^{3}}{c^{3}}\Omega^{2}\alpha
\frac{\bm{M}}{M}\cos^{2}{\chi},
  \label{6}
\end{eqnarray}
where $\alpha$ is a parameter characterizing the magnitude of
the electric current flowing through the pulsar tube.
If a pulsar spends its energy for the generation and
acceleration of plasma in the magnetosphere, and the star spins down due to the
currents flowing in a magnetosphere and closing on the star surface, then the angle
$\chi$ approaches to 90$^{\circ}$. This model gives an interesting relation of
$\nu\sin{\chi}$=const, where $\nu=\Omega/2\pi$ is the rotation frequency of the star.
This relation requires that a pulsar stops spinning down, instead of rotates at a
constant rotation frequency, when its inclination angle $\chi$ increases to
$90^{\circ}$. This appears to contradict observations.

\subsubsection{Complex braking models}
 A pulsar's secular spin-down could be simultaneously caused by several different energy loss mechanisms such as magnetic dipole radiation, particle wind, star-disk interaction, neutrino
emission and gravitational radiation. The loss of rotational
energy of a pulsar is accompanied by the change of $\chi$.
Taking a complex braking model composed of the magnetic-dipole
and current mechanisms for example, the total braking torque applied to a pulsar is
$\bm{K}=\bm{K}_{\text{dip}}+\bm{K}_{\text{cur}}$, then the torque equations become\,\citep{barsukov2009ARep}
\begin{eqnarray}
    ~~~I\frac{d\Omega}{dt}=-\frac{2}{3}\frac{M^{2}}{c^{3}}\Omega^{3}(\sin^{2}{\chi}+\alpha \cos^{2}{\chi}),
  \label{7}
\end{eqnarray}
\begin{eqnarray}
    I\frac{d\chi}{dt}=\frac{2}{3}\frac{M^{2}}{c^{3}}\Omega^{2}(\alpha-1)\sin{\chi}\cos{\chi}.
     \label{8}
\end{eqnarray}
Here $\alpha$ changes over time, if $(\alpha-1)>0$, then $\dot{\chi}>0$, while $(\alpha-1)<0$ and $\dot{\chi}<0$.
\subsection{Internal origin}
\subsubsection{Cooling and dipole-dipole  interaction}
The internal origin of the inclination angle change is related to the internal physics of neutron stars. Utilizing a late-stage proto-neutron star model with
a strong toroidal field,~\citet{Lander2018} found that the evolution of $\chi$ is influenced
by a combination of neutrino cooling, external electromagnetic torques, and
internal dissipation. They show that, before the formation of the neutron star shell and before the neutron superfluidity and proton superconductivity internal dissipation will act to make $\chi$ increase for star whose magnetic field-induced
deformation is prolate. Their results show that a typical pulsar evolves into an near-orthogonal
rotators and the millisecond magnetar with a small initial magnetic inclination and
with a magnetic field greater than $5\times10^{15}$\,G evolves into near-aligned rotators.
\begin{figure}[t]\centering
\centerline{\includegraphics[width=0.9\linewidth]{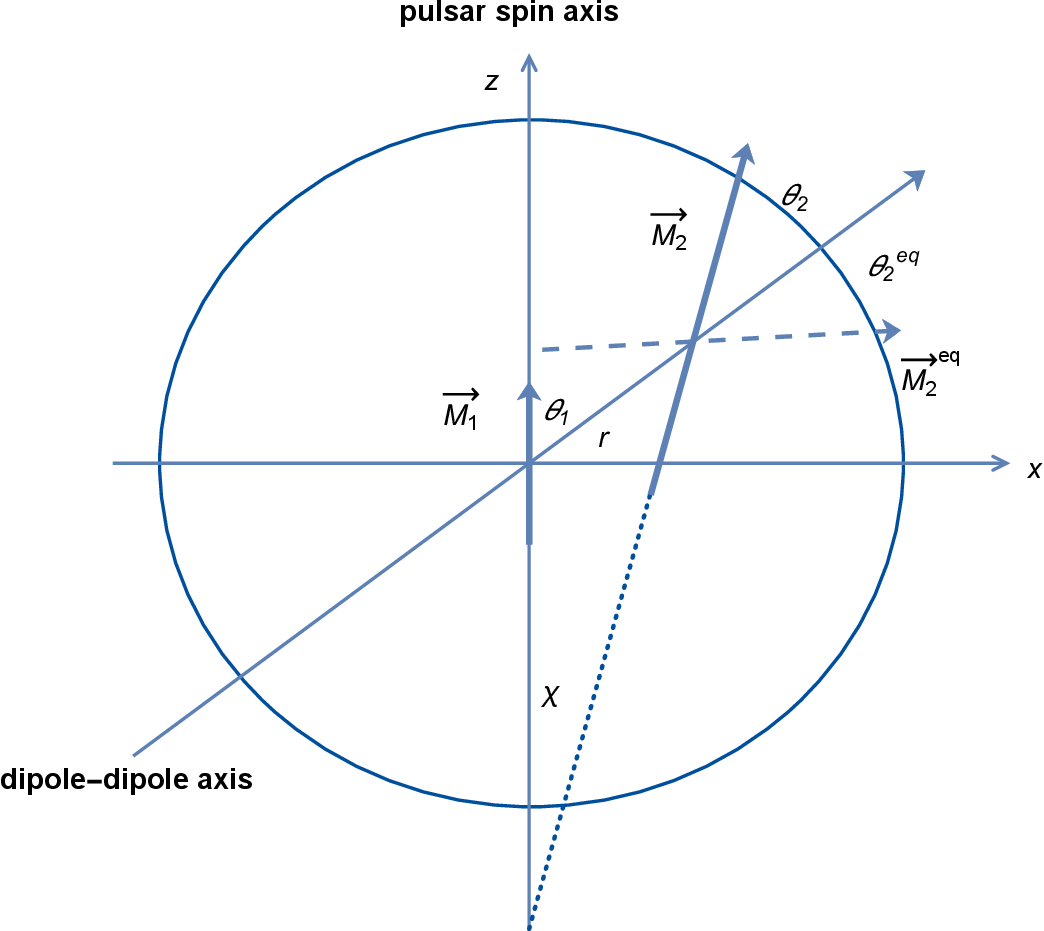}}
\vspace{-0.8 em}	
 \caption{The two-dipole system in the pulsar. $\Vec{M_{1}}$ is fixed and aligned with the spin axis $z$. The dipole-dipole axis $r$ makes an angle $\theta_{1}$with $z$. The dashed line corresponds to $\Vec{M_{2}}$ in the equilibrium position. For more details see~\cite{Hamil2016}. \label{fig1}}
\vspace{-1.6 em}
\end{figure}

~\citet{Hamil2016} explored a double magnetic-dipole model of pulsars with a constant moment
of inertia and magnetic dipole moment but variable magnetic inclination angle. In this model (Figure \ref{fig1}), there are two magnetic moments $M_{1}$ and $M_{2}$ inside the star. The former is generated by the rotation effect of a charged sphere, and the latter is generated by the magnetization of ferromagnetically ordered material. There will be mutual attraction or repulsion between $M_{1}$ and $M_{2}$, resulting in a change in
$\chi$ for a pulsar. Whether an inclination angle increases or decreases and its change rate are dependent
on a number of factors, such as the strength of the two magnetic moments, the minimum energy position.

It is proposed that the double magnetic-dipole model for pulsar inclination angle
evolution may be most promising because it not only naturally explains
the origin of magnetic fields in young and strongly magnetized pulsars,
but also can account for some observable phenomena in pulsars, such as
profile shifting, see the next section for details.
\subsubsection{Dissipation and precession }
Let's consider a rotating pulsar in Cartesian coordinates. It is assumed that the principal symmetry axis of the pulsar is along the $zz$ direction. Due to the centrifugal force effect, the pulsar should be oblate rotating along the $zz$ axis. If the matter density distribution of the pulsar is asymmetric, the $zz$ axis and the spin axis no longer coincide, and there is an angle between them, denoted as $\beta$. The value of $\beta$ is affected by many factors, such as the rate of spin down and so on. As pulsar spins down, its rotational angular velocity also precessions around the $zz$ axis, and the precession process affects the evolution of the inclination angle, including the magnitude and direction of evolution. Describing the above process in detail relies on the Euler's equations. It is worth emphasizing that in a pulsar precession cycle, the increase and decrease of the $\chi$ appear alternately. When $\chi$ is close to $0^{\circ}$, the change is very slow, but there will not be zero degrees. Similar change will occur in the inclination angle increase phase as $\chi$ is close to $90^{\circ}$.

For example, considering the combined effect of magnetospheric current and magnetic dipole radiation~\citet{barsukov2009ARep} gave the precession period that is
\begin{eqnarray}
    T_{\text{prec}}=\frac{2\pi}{\Omega_{\text{prec}}},~ \Omega_{\text{prec}}=-\frac{3}{5}\Omega\frac{M^{2}}{Rc^{2}I}\cos{\chi}.
 \label{9}
\end{eqnarray}

~\citet{Dall2017} studied the effect of viscous dissipation on the magnetic inclination
distribution of young pulsars. The evolution equation for the inclination angle is 
\begin{eqnarray}
 \tau_{\text{d}}\equiv \frac{E_{\text{prec}}}{|\dot{E}_{\text{diss}}|},~~~~~
\frac{d\chi}{dt}=\frac{\cos{\chi}}{\sin{\chi}\tau_{d}}, 
\label{10}
\end{eqnarray}
where $\tau_{\text{d}}$ is timescale for dissipation of the freebody precession, $E_{\text{prec}}$ the precession energy and $\dot{E}_{\text{prec}}$ the dissipation rate.
They showed that the dissipation of precessional
motions by bulk viscosity can naturally produce a bi-modal distribution of tilt angles,
and proton superconductivity reduces the $\beta$-reaction rate, which causes the volumetric
viscosity coefficient and the neutrino cooling rate to drop. Applied to the Crab pulsar, they
found that there may be additional viscous processes, such as crust-core coupling
via mutual friction in the superfluid core, which may affect the evolution of magnetic
inclination angle over longer timescales.

~\citet{Kraav2017} considered the evolution of inclination angle and precession damping
of radio pulsars and assumed that the neutron star consists of 3 "freely" rotating components:
the crust and two core components, one of which contains pinned superfluid vortices, Within
the framework of this model the star simultaneously can have glitch-like events combined
with long-period precession (with periods $10-10^{4}$ yrs). They found that the case of the
small quantity of pinned superfluid vortices seems to be more consistent with observations.
\section{On the trends and forms of the inclination angle evolution}\label{sec3}
\subsection{The evolution trends}
Increasing or decreasing of the $\chi$ depends on the mechanism of a pulsar energy
losses. There are three main variation trends for the pulsar inclination angle: near-aligned,
near-orthogonal and oscillation trends.
\begin{enumerate}
\item For the first evolution trend, the inclination angle $\chi$ decreases either when
$\chi$ is less than $\pi/2$ and the rotation and magnetic
axes are moving toward alignment, or when $\chi$ is greater than $\pi/2$ and
the rotation and magnetic axes are counter-aligning.
\item  For the second variation trend, although both internal dissipation and magnetospheric current models can increase the magnetic inclination angle, there is a difference. The internal dissipation that is temperature dependent dominate the early evolution of $\chi$ in pulsars and gradually give way to electromagnetic torques as the neutrino cooling process progresses.
The magnetospheric current model can cause a long-term increase in magnetic inclination angle.
\item For the third variation trend, similar to a simple pendulum, the inclination angle $\chi$ of a pulsar is undergoing a long and permanent oscillation\,($\chi$ increases $\rightarrow$ decreases $\rightarrow$ increases $\ldots$). If a neutron star is triaxial ellipsoid, the angle $\chi$ permanently oscillates around an equilibrium angle because of precession\,\citep{melatos2000MN}. Also the oscillation trend has been predicted in the double magnetic-dipole model\,\citep{Hamil2016}.
\end{enumerate}
\subsection{The evolution forms}
\begin{figure}[t]
\centerline{\includegraphics[width=1\linewidth]{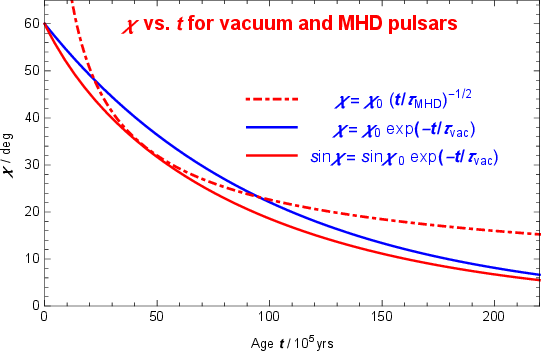}}
\vspace{-0.8 em}	
 \caption{Plot of magnetic inclination evolution comparison under vacuum model (solid lines) and
plasma-filled magnetosphere model (dotted line). We arbitrarily choose parameters: $\chi_{0}=60^{\circ}$;
$\tau_{\text{MHD}}=1.42\times10^{6}$yrs; $\tau_{\text{vac}}=1.0\times10^{7}$yrs.
\label{fig2}}
\vspace{-1.6 em}
\end{figure}
There is also statistical evidence that the inclination angle of pulsars tends to achieve alignment in the long term~\citep{Yong2010MNRAS,Lyne1988MNRAS}.
Increasing inclination angle of Crab pulsar may require a further ingredient such as the presence of return currents in the magnetosphere or precession\,\citep{Zanazzi2015}. But an orthogonal rotator with plasma-filled force-free magnetosphere\,\citep{Philippov2014} requires a much larger current  
than used in~\citet{beskin2007ApSS}, so that the minimum spin-down energy losses correspond to that of an aligned rotator in the end.

According to vacuum model,  a pulsar would stop spinning down when
alignment is achieved in obvious contradiction with observations. It was suggested by~\citet{Goldreich1969} that the progress of alignment would be slowed down by dissipative processes for a non-spherical pulsar.
The magnetic inclination evolution in vacuum model is exponential,
\begin{eqnarray}
    \chi=\chi_{0}\exp{(-t/\tau_{\text{vac}})},
\label{11}
\end{eqnarray}
or
\begin{eqnarray}
    \sin{\chi}=\sin{\chi_{0}}\exp{(-t/\tau_{\text{vac}})},
\label{12}
\end{eqnarray}
where $\tau_{\text{vac}}$ is the alignment time-scale of a vacuum pulsar. The sinusoidal form of
Eq.(12) is convenient for calculating torques.
After considering that plasma fills the magnetosphere, the evolution of $\chi$ will be described by a relatively slow nonlinear form,
\begin{eqnarray}
    \chi=\chi_{0}(t/\tau_{\text{MHD}})^{-1/2},
\label{13}
\end{eqnarray}
where $\tau_{\text{MHD}}$ is the MHD pulsar alignment time-scale.

In figure \ref{fig2}, we present a sketch that magnetic inclination angle evolves over time
for the vacuum case and the MHD case. The inclination angle of MHD case
diverges when time approaches zero. Although the plasma-filled magnetosphere model 
is relatively close to the realistic pulsar, its complex magnetospheric structure brings greater uncertainty.
In this paper, we will choose simple vacuum model to study the inclination angle evolution for 12 pulsars from the Nashan observations, see the next section for details.
\begin{table*}[t]
\centering
\caption{The spin frequencies, their derivatives, inclination angles and their change
rates for pulsars with known braking indices. The superscripts 1 and 2 denote that the values of $\dot{\chi}$ in Columns 8 and 9 are obtained using vacuum and
MHD models, respectively. References: [1]~\citet{Lyne1993};[2]~\citet{Lyne1996};
[3]~\citet{Livingstone2007};[4]~\citet{Espinoza2011};[5]~\citet{Roy2012MNRAS};
[6]~\citet{Wel2011MNRAS};[7]~\citet{Harding2008ApJ};[8]~\citet{Dyks2003ApJ};
[9]~\citet{Watters2009ApJ};[10]~\citet{Li2013ApJL};[11]~\cite{Takata2007ApJ};
[12]~\citet{Zhang2000};[13]~\citet{Nikitina2017ARep};[14]~\citet{Wang2014MNRAS};
[15]~\citet{Rookyard2015MNRAS}. Similar calculations of $\dot{\chi}_{1}$ were performed by~\citet{tian2018NewA}\label{tab1}}
\vspace{-1 em}
\scalebox{1.6}{
\small
\begin{tabular}{lcccccccc}
\bottomrule
Pulsar & $\nu$  & $\dot{\nu}$  & $n$ & Refs & $\chi$ & Refs& $\dot{\chi}_{1}$
& $\dot{\chi}_{2}$\\
 & (s$^{-1}$)  & ($10^{-11}$s$^{-2}$) & & & ($^{\circ}$) & & ($^{\circ}$/100yr) &($^{\circ}$/100yr)\\
\midrule
Crab & 30.2254  & -38.6228  &  2.51(1)  &[1] & 45 & [7] &0.558 & 1.412 \\
&   &   &    & & 60 & [8] & 0.966& 1.622 \\
&   &   &    & & 70 & [9] &8.486 & 1.877\\
Vela & 11.2     & -1.57 &  1.4(2) & [2] & 62-68  &[9] & 0.38-0.51 & 0.82-0.98 \\
    &      &  &   &  &77-79  & [9] & 0.87-1.04 & 1.66-1.94 \\
J1833-1034     & 16.1594  & -5.2751  & 1.857(1)&[5] & 70 &[10] & 0.926 &1.727 \\
J0540-6919    &19.8345   & -18.8384  & 2.140(9) &[3] & 30 &[11] & 0.425& 1.865\\
          & &  &  & & 50 & [12] &0.877 &2.16 \\
J1734-3333     & 0.8552  & -0.1667  & 0.9(2) &[4] & 21 & [13] & 0.142 & 1.247   \\
J1846-0258     & 3.0782 & -6.7156  & 2.65(1) & [3] &  10 &[14] & 0.121 & 4.12\\
J1119-6127     & 0.408  & 40.2  & 2.684(2)  &[6] & 21 & [15] &0.108 &0.949 \\
 J1513-5908     & 0.151  & 15.3  & 2.837(1)  &[3]&60 &[12] & 0.255& 0.595\\
     & &  &   & & 30&[15] & 0.471 & \\
\bottomrule
\end{tabular}}
\vspace{-1.5 em}
\end{table*}

\section{On the observed effect of the evolution of Inclination angle}\label{sec4}
\subsection{Slow glitches}
Pulsar glitches are rare events of very short duration, shown as sudden jumps 
in rotational frequency\,\citep{yuan2017MNRAS}. The slow glitch was first coined 
when\,\citep{zou2004MN} tracked the evolution of the frequency $\nu$ and first-order 
frequency derivative $\dot{\nu}$ of PSR B1822–09, with a continuous increase in $\nu$ 
over several hundreds of days. This process corresponds to an impulsive decrease in 
the spin-down rate followed by an exponential increase to its pre-glitch value.
Three slow glitches were observed in PSR B1822-09 by\,\citep{istomin2007ARep}. 
They explained the slow glitches well in a current-loss model in which the star’s shape is 
rearranged during the inclination angle evolution because of the
decelerating torque acting on the star. An orthogonal rotator fits the proposed model 
better than a co-axial rotator, which means that the inclination angle of the
radio pulsar tends to 90$^{\circ}$ during its evolution. 
\subsection{Radiation pulse profile}
The inclination angle is an important parameter in the geometry of pulsar radiation that has a significant impact on changes in the pulse profile. \cite{Rankin1990ApJ} reported that the core-component widths $W_{\text{core}}$ depend only upon the pulsar period $P$ and inclination angle $\chi$, 
\begin{eqnarray}
    W_{\text{core}}=2.45^{\circ}P^{-1/2}/\sin{\chi}.
    \label{14}
\end{eqnarray}
~\citet{krzysztof2011MN} examined the validity of Eq.(14) by performing a statistical analysis of half-power pulse–widths of the core components in average pulsar profiles.~\citet{tedila2022ApJ} estimated the inclination angle of a long period pulsar (PSR J1900+4221) by using  the Eq.(14) and obtained $\chi=7^{\circ}$. 

The nulling fraction\,(NF) was proposed in positive correlation with the age and the rotation period of a pulsar\,\citep{Biggs1992ApJ}. The evolution of the pulsar age is suggested in correlation with the inclination angle $\chi$. From consideration of energy loss through magnetic dipole radiation, the evolution of $\chi$ is from large to small. However, consideration of the longitudinal current flow and the pair production in the magnetosphere~\citet{beskin1988ApSS} suggests that the change of $\chi$ is from small to large as a pulsar ages. 
\subsection{Pulsar braking indice}
A pulsar spins down due to electromagnetic radiation, particle winds, neutrino emission, or gravitational
radiation. The long-term slowdown of the star follows a power-law form $\tau_{ext}\propto\Omega^{n}$, 
where $\tau_{ext}$ is the external torque acting on the crust, and $n$ is the braking index, which is an 
important quantity relating to the energy loss mechanism and determined by observations,
\begin{eqnarray}
n=\frac{\nu\ddot{\nu}}{\dot{\nu}^2}
=2-\frac{P \ddot{P}}{\dot{P}^2},
\label{15}
\end{eqnarray}
where $\ddot{\nu}$ is the second derivative of $\nu$, and $\dot{\nu}$ is the derivative of $\nu$
It is generally assumed that the moment of inertia and the magnetic dipole moment are constant. If all the rotational energy loss of the pulsar is converted into magnetic dipole radiation, using Eq.(1), we get
\begin{eqnarray}
n=3+2\frac{\nu}{\dot{\nu}}\frac{\dot{\chi}}{\tan{\chi}}.
\label{16}
\end{eqnarray}
Taking into account the plasma effect, equation (16) become
\begin{eqnarray}
    n=3+2\frac{\nu}{\dot{\nu}}\frac{\dot{\chi}\sin{\chi}\cos{\chi}}{1+\sin^{2}{\chi}}.
    \label{17}
\end{eqnarray}
From Eqs(16) and (17), $\dot{\nu}<0$, when $\dot{\chi}>0$,
there is $n<3$, whereas $\dot{\chi}<0$, then $n>3$. Table 1 shows eight young pulsars with known braking indice. For comparison, we calculate the rate of change in magnetic
inclination for the vacuum case and the MHD case by using Eq.(16) and Eq.(17), respectively.

For decades, despite their great success both vacuum and plasma-filled models should be combined with other radiation mechanisms such as gravitational wave, pulsar wind and internal dissipation to simulate the magnetosphere current, rotating, cooling and magnetic field evolution of pulsars.

A combination of gravitational wave and magnetic energy loss mechanisms radiation could give rise to a braking index $3<n<5$\,\citep{Araujo2016}. To account for the observed braking indices, several other interpretations have been put forward, see~\citet{Gao2017} for a brief summary.

Table 1 does not include high-braking index PSR J1640-4631, whose value of $\chi$ is predicted by the plasma-filled magnetosphere model\,~\citet{Eksi2016}. Based on the estimated ages of their potentially associated supernova remnants\,(SNRs) and the timing parameters,~\citet{Gao2016} measured the mean braking indices of eight magnetars with SNRs, and interpret the braking indices of $n<3$ within a combination of MDR and wind-aided braking, while the larger braking indices of $n>3$ for other three magnetars are attributed to the decay of external braking torque, which might be caused by magnetic field decay.

\begin{table*}[ht]
\captionof{table}{Rotation and magnetic inclination angle evolution parameters ($\nu$,
$\dot{\nu}$, $\ddot{\nu}$, $n$, $\tau_{c}$, $\chi$ and $\dot{\chi}$) of 12 pulsars without glitches. The numbers in the parentheses represent the calculation errors. The data in Columns 1-6 and 10 are from~\citet{Dang2020}.}
\vspace{-1 em}
\scalebox{0.95}{
\begin{tabular}{lccccccccc}
	\bottomrule
	Pulsar name& Epoch & $\nu$  & $\dot{\nu}$  & $\ddot{\nu}$ & $n$ & $\tau_{c}$ & $\chi$ &
$\dot{\chi}$ & Data Range\\
	  (J2000)   &    & (s$^{-1}$)&($10^{-14}$s$^{-2}$) & ($10^{-25}$s$^{-3}$)& & ($10^{5}$ yrs)
& ($^{\circ}$) & ($^{\circ}$/100yrs)& (MJD)  \\
	\midrule
	J0157+6212 & 54594 & 0.425201 & -3.414773(2) & 0.136(7) & 4.8(2)  & 1.974 & 46.5(1.5) & -0.0137(11)  & 52470–56719 \\
	J0614+2229 & 54595 & 2.985206 & -5.27267(1)  & 53.8(6)  & 56.6(6) & 8.978 & 11.93(6)  & -0.0165(1) & 52473–56718\\
    J1739-2903 & 54598 & 3.097068 & -7.554400(5) & 0.13(2)  & 58.8(8) & 6.501 & 11.72(8)  & -0.0233(2) & 52495–56702\\
	J1743-3150 & 54599 & 0.414143 & -2.071559(2) & 0.067(5) & 6.5(7)  & 3.171 & 37(8)     & -0.012(1)   & 52495–56702\\
    J1759-2922 & 54589 & 1.740942 & -1.40276(1)  & 0.07(2)  & 33(19)  & 1.968 & 15.6(1.2) &-0.0056(26) & 52496–56683\\
	J1820-1346 & 54607 & 1.085232 & -0.528865(3)  & 0.015(9) & 60(34)  & 3.254 & 12(4)     &-0.0047(12) & 52496–56719\\
    J1833-0338 & 54603 & 1.456193 & -8.815136(2) & 1.71(6)  & 34(1)   & 2.619 & 14.3(2)   &-0.043(1)   & 52496–56719\\
    J1857+0526 & 54619 & 2.857527 & -5.658594(4) & 0.19(1)  & 10(1)   & 8.008 & 28(2)     &-0.0067(5)  & 52520–56718\\
	J1902+0556 & 54288 & 1.339437 & -2.309011(5) & 0.09(2)  & 23(5)   & 9.199 & 18(2)     &-0.0098(13) & 52473–56103\\
    J1916+0951 & 54587 & 3.700203 & -3.448575(6) & 0.10(2)  & 23(6)   & 1.702 & 17(3)     &-0.0053(7)  & 52473–56700\\
	J1918+1444 & 54644 & 0.846658 & -15.20376(3) & 6.39(8)  & 22.9(3) & 8.831 & 17.6(1)   &-0.102(1)   & 52569–56719\\
    J2004+3137 & 54539 & 0.473643 & -1.671669(2) & 0.338(6) & 62.6(8) & 4.393 & 10.38(7)  &-0.0348(3)   & 52488–56590\\
	\bottomrule
\end{tabular}}
	\vspace{-1.5 em}
\end{table*}
Very recently,~\citet{Yan2021} have applied a two-dipole model to two high-n magnetars SGR 0501+4516 and IE 2259+586, and estimated their initial magnetic moments, initial inclination angles and average decrease rates of inclination angles. A comparisons with rotationally powered pulsars are presented. The decreasing magnetic inclination angle and narrow pulse width of a magnetar support its lack of radio emission.

\section{An application of inclination angle evolution in vacuum model}\label{sec4}

\subsection{Model and data}
In this section, using an alignment rotator model in vacuum, we investigate the magnetic inclination angle change rates for 12 high-braking index pulsars without glitch, whose timing observations are obtained using the Nanshan 25-m Radio Telescope at Xinjiang Astronomical Observatory. Our data come from~\citet{Dang2020}. Although the plasma-filled magnetosphere model is relatively close to the realistic pulsar, its complex magnetospheric structure brings greater uncertainty.In this paper, we choose a relatively simple vacuum model to study the rotation evolution of pulsars. From Eqs(1) and (2), we get
\begin{eqnarray}
    \dot{\chi}=\frac{\dot{\nu}}{\nu}\cot{\chi}.
    \label{18}
\end{eqnarray}
From Eq.(18), one can get that the projection of the rotation angle
frequency on the rotation axis is a constant,
\begin{eqnarray}
\nu \cos{\chi}=const.
\label{19}
\end{eqnarray}
Since $\nu$ always decreases, this requires a increasing $\cos{\chi}$ and thus a decreasing $\chi $, ensuring the product of $\nu \cos{\chi}$ is a constant.
Combined with Eq.(16) we have
\begin{eqnarray}
n=3+2\cot^{2}{\chi}.
\label{20}
\end{eqnarray}
The second term on the right of Eq.(20) is positive, so $n>3$. Then we get,
\begin{eqnarray}
    \cot{\chi}=(\frac{n-3}{2})^{\frac{1}{2}}.
    \label{21}
\end{eqnarray}
\begin{figure}[t]\centering
\centerline{\includegraphics[width=1\linewidth]{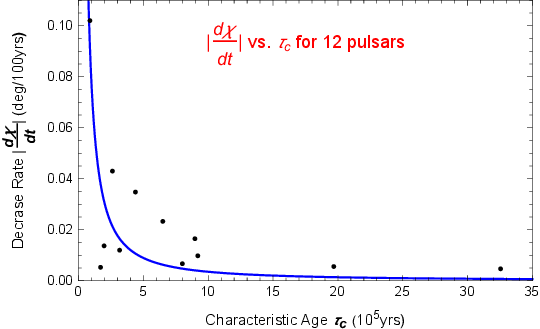}}
\vspace{-0.8 em}	
 \caption{The rate of change in inclination angle as a function of
characteristic age. 
\label{fig3}}
\vspace{-1.6 em}
\end{figure}
~\citet{Dang2020} used the Nanshan 25-m radio telescope of Xinjiang Observatory to
obtain the observation data of 87 pulsars for 12 years. These pulsars do not have
glitches during the observation period. We selected 12 pulsars with the braking indice
$3<n<100$. Table 2 shows some parameters of these pulsars including inclination angles and their change rates calculated by using Eq.(18) and Eq.(21).

Then we use the data in Table 2 to fit the relationship between the change rates $\dot{\chi}$ and the age $\tau_{c}$. Figure 3 shows the fitting results.
\begin{figure*}[t]\centering
\centerline{\includegraphics[width=1\linewidth]{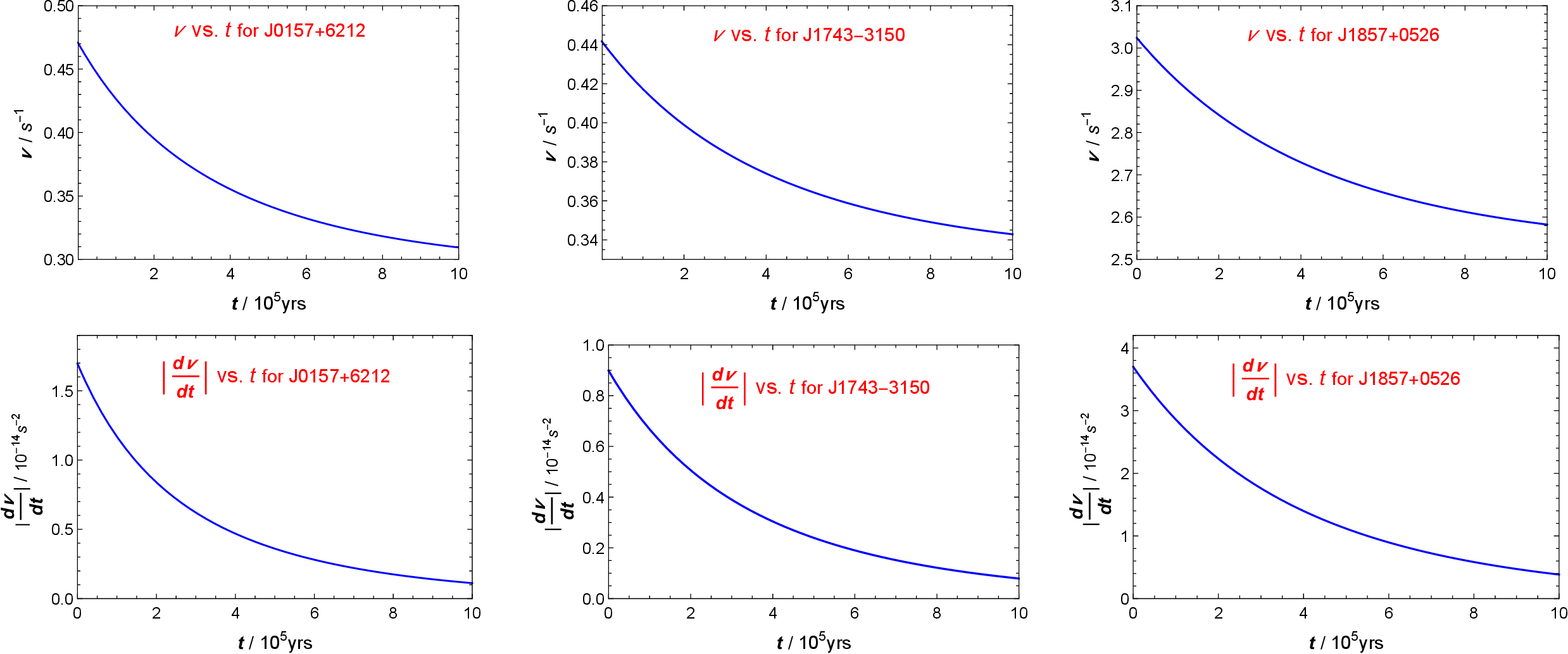}}
\vspace{-0.8 em}	
 \caption{Evolutions of $\nu$ and $\dot{\nu}$ over time $t$ for J0157+6212, J1743-3150 and J1857+0526. The timescale of the evolution of magnetic 
inclination angle is taken as $\tau_{A}=10^{6}$yrs for three sources. 
\label{fig4}}
\vspace{-1.3 em}
\end{figure*}
In the fitting formula (22), $\dot{\chi}$ is in units of deg$/100$yrs, $\tau_{c}$  is in units of $10^{5}$yrs,
\begin{eqnarray}
|\dot{\chi}|=a\times\tau_{c}^{-b},
\label{22}
\end{eqnarray}
and the fitting parameters are $a=0.77(13)$ and $b=1.34(37)$, where the numbers in the parentheses represent the errors. As can be seen from Figure \ref{fig3}, as pulsars spin down, the change rate of inclination angle $\dot{\chi}$ declines continuously, the decline is faster in the early evolution and slower in the late evolution.

\subsection{Three representative pulsars}
In pulsar observation, due to the influence of timing noise, the observed data will deviate
from the actual situation, which will affect the measurement of certain parameters. Among
these 12 sources, we believe that the source with a relatively large braking index has a large deviation from the actual one, and such
a source needs to be further observed in the future. So we choose three pulsars J0157+6212,
J1743-3150, J1857+0526 with $3<n<10$ for further study.

~\citet{Philippov2014} shows that the vacuum pulsars come into alignment exponentially.
To explore the evolution of $\nu$ over time, we use the data in Table 2 to obtain the constant
of motion (Table 3) and assume that the evolution of $\chi$ within the vacuum model
is exponential as described by Eq.(11).
\begin{table}[t]\centering
\captionof{table}{Fitted parameters of three pulsars with braking index $3<n<10$ under the
vacuum model. The alignment time-scale is taken as $10^{6}$yrs. }
\vspace{-1.5 em}
\scalebox{1.7}{
\small
\begin{tabular}{lcccc}
	\bottomrule
	Pulsar  &    $t_{a}$  & $\chi_{0}$  & $\nu_{0}$  & $const.$ \\
    (J2000) & ($10^{5}$yrs) &($^{\circ}$) & (s$^{-1}$) &          \\
	\midrule
	J0157+6212 & 1.039 & 51.60 & 0.47 & 0.293  \\
	J1743-3150 & 1.153 & 41.52 & 0.44 & 0.331 \\
    J1857+0526 & 1.779 & 33.45 & 3.02 & 2.523 \\
	\bottomrule
\end{tabular}}
	\vspace{-1.5 em}
\end{table}
The real ages of these three pulsars are estimated as  $t_{a}\approx -\frac{1}{n-1}\frac{\Omega}{\dot{\Omega}}=\frac{1}{n-1}
\frac{P}{\dot{P}}$. The spindown rate is then given as
\begin{eqnarray}
    \dot{\nu}=\frac{\sin{\chi}}{\cos^{2}{\chi}}\dot{\chi} const.
    \label{23}
\end{eqnarray}
Using Eqs(19) and (23), we make plots of $\nu$ and $\dot{\nu}$ over $t$ for pulsars J0157+6212, J1743-3150 and J1857+0526, shown as in Figure \ref{fig4}.
Table 3 lists the fitting parameters for the three pulsars including the actual
age $t_{a}$, initial magnetic inclination $\chi_{0}$ and the initial frequency $\nu_{0}$.

The possible relation between the evolution of the magnetic inclination 
and the rotation frequency has been investigated by~\citet{tian2018NewA}.
It is assumption that the various timing behavior unexpected by the magnetic dipole 
radiation is due to a secular change of $\chi$. Many other effects can also result 
in the same timing behavior such as the evolution of the magnetic field and an
interaction between fall-back disk and magnetic field\,\citep{chen2016MN}. If we take into consideration these effects, the actual evolution of
the magnetic inclination angles in Table 2  might differ very much from the values calculated.

\section{SUMMARY}\label{sec5}
In this paper, we have performed a short review on the the pulsar inclination angles, as well as their evolutions and given an application of inclination angle evolution in vacuum to 12 high-$n$ pulsars from the Nashan observation. In the future, it is expected that polarization observations will give more values of the pulsars' inclination angles and their variations. We will combine observation phenomena to constrain the theoretical models, and explore the actual evolution process of these sources as much as possible.

\section*{Acknowledgments}
This work was supported by Chinese National Science Foundation through No.12041304.

\bibliography{IWARA}

\end{document}